\documentclass{article}

\newcommand{\ml}{l\kern-0.035cm\char39\kern-0.03cm}

\newtheorem{prop}{Proposition 1.}

\newtheorem{defi}{Definition 1.}

\newtheorem{prop2}{Proposition 2.}

\newtheorem{defi2}{Definition 2.}

\newtheorem{ex2}{Example 2.}

\begin{document}

\title{Representation theorem for obsevables on a quantum logic}
\author{Andrei Khrenikov\\
International Centre for Mathematical Modeling\\
in Physics and Cognitive Sciences,\\
MSI, University of V\"axj\"o, S-35195 Sweden\\
E-mail:Andrei.Khrennikov@msi.vxu.se\\
Olga N\'an\'asiov\'a\\
Department of Mathematics and Descriptive Geometry\\
Faculty of Civil Engineering\\
Slovak University of Technology\\
Radlinsk\'eho 11,
813 68 Bratislava,Slovakia\\
Email: olga@res.svf.stuba.sk}
\maketitle
\date{}

\begin{abstract}
We will study a conditional state on a quantum logic using
Renyi's approach (or Baye\-sian principle). This approach helps us to define
independence of events and differently from the situation in the classical
theory of probability, if an event $a$ is independent of an event $b$, then
the event $b$ can be dependent on the event $a$. 
We will show that we can define a $s$-map (function for
simultaneous measurements on a quantum logic). It can be shown that if we
have the conditional state we can define the $s$-map and conversely. By
using the $s$-map we can introduce joint distribution also for noncompatible
observables on a quantum logic.
\end{abstract}

\section{Introduction}

The idea that quantum probabilities are nothing other than
conditional probabilities was intesively discussed from various points
of views, see, e.g., \cite{A}--\cite{DEM}. Recently this approach to quantum probability
was generalized in the so called  contextual framework \cite{KHR1}--\cite{KHR6}. 
In the latter
approach we consider conditioning with respect to various contexts, complexes
of experimental physical conditions and not conditioning of one event, $a,$ 
with respect to other event, $b,$  under the same context (state) as it was
done
in investigations on conditional probabilities (compare with Kolmogorov \cite{K1}, \cite{K2},
Renyi \cite{Ren}, Cox \cite{Cox},  
or  quantum logic generalizations \cite{Bel}, see also \cite{A}--\cite{DEM}).

In the present paper we still use the traditional event-conditioning (under
the
fixed conditional state). However, we essentially generalize the notion of
a conditional state (here we continue investigations \cite{N2}--\cite{N7}).

We will study a conditional state on a quantum logic using
Renyi's approach (or Baye\-sian principle). This approach helps us to define
independence of events and differently from the situation in the classical
theory of probability, if an event $a$ is independent of an event $b$, then
the event $b$ can be dependent on the event $a$ (problem of causality) (\cite
{N6}, \cite{N7}). We will show that we can define a $s$-map (function for
simultaneous measurements on a quantum logic). It can be shown that if we
have the conditional state we can define the $s$-map and conversely. By
using the $s$-map we can introduce joint distribution also for noncompatible
observables on a quantum logic. Moreover, if $x$ is an obsevable on $L$ and $%
B$ is Boolean sub-algebra of $L$, we can construct an observable $z=Exp(x\vert
B)$, which is a version of conditional expectation of $x$ but it need not to
be necessarily compatible with $x$.

\vskip 2pc \centerline{\bf {1. A conditional state on a quantum logic } }
\vskip 2pc In this part we introduce the notions as a quantum logic, a
state, a conditional state and their basic properties.

\begin{defi}
Let $L$ be a nonempty set endowed with a partial ordering $\leq $. Let there
exists the greatest element ($1$) and the smallest element ($0$). Let there
be defined the operations supremum $(\vee )$, infimum $\wedge $ (the lattice
operations ) and a map $\perp :L\to L$ with the following properties:

\begin{itemize}
\item[(i)]  For any $\{a_n\}_{n\in\mathcal{A}}\in L$, where $\mathcal{%
A\subset N}$ are finite 
\[
\bigvee_{n\in\mathcal{A}} a_n, \bigwedge_{n\in\mathcal{A}} a_n\in L. 
\]

\item[(ii)]  For any $a\in L$ $(a^\bot )^\bot =a$.

\item[(iii)]  If $a\in L$, then $a\vee a^\bot =1$.

\item[(iv)]  If $a,b\in L$ such that $a\le b$, then $b^\bot\le a^\bot $.

\item[(v)]  If $a,b\in L$ such that $a\le b$ then $b=a\vee (a^\bot\wedge b)$
(orthomodular law).
\end{itemize}

Then $(L,0,1,\vee ,\wedge ,\perp )$ is called \textit{the quantum logic}
(briefly \textit{q.l.}).
\end{defi}

\vskip 1pc Let $L$ be q.l.. Then elements $a,b\in L$ will be called:

\begin{itemize}
\item  \textit{orthogonal} ($a\bot b$) iff $a\le b^\bot$;

\item  \textit{compatible} ($a\leftrightarrow b$) iff there exist mutually
orthogonal elements $a_1,b_1,c\in L$ such that 
\[
a=a_1\vee c\mbox {\hskip 1pc  and\hskip 1pc } b=b_1\vee c. 
\]
\end{itemize}

\vskip 1pc If $a_i\in L$ for any $i=1,2,3,...$ and $b\in L$ is such, that $%
b\leftrightarrow a_i $ for all $i$, then $b\leftrightarrow\bigvee_{i=1}^n
a_i $ and 
\[
b\wedge (\bigvee_{i=1}^\infty a_i)=\bigvee_{i=1}^\infty ( a_i\wedge b) 
\]
(\cite{D2},\cite{P1},\cite{V1}). \vskip 1pc A subset $L_0\subseteq L$ is 
\textit{a sub-logic of $L$ } if for any $a\in L_0$ we have $a^\bot\in L_0$
and for any $a,b\in L_0$ $a\vee b\in L_0$. \vskip 1pc

\begin{defi}
A map $m:L\to [0,\infty ) $ such that

\begin{itemize}
\item[(i)]  $m(0)=0$ and $m(1)=1$.

\item[(ii)]  If $a\bot b$ then $m(a\vee b)=m(a)+m(b)$
\end{itemize}

is called \textit{a state} on $L$. If we have orthomodular $\sigma $-lattice
and $m$ is $\sigma $-additive function, then $m$ will be called a \textit{$%
\sigma $-state}.
\end{defi}

\vskip 1pc

\begin{defi}
\cite{N7} Let $L$ be a quantum logic. A subset $L_c\subset L-\{0\} $ is
called \textit{a conditional system (CS) in $L$} \textit{($\sigma$-CS in $L$ 
})if the following conditions hold:

\begin{itemize}
\item  If $a,b\in L_c$, then $a\vee b\in L_c$. (If $a_n\in L_c$, for $%
n=1,2,...$, then $\bigvee_n a_n\in L_c$.)

\item  If $a,b\in L_c$ and $a< b$, then $a^\perp\wedge b\in L_c$.
\end{itemize}
\end{defi}

\vskip 1pc Let $A\subset L$. Then $L_c(A)$ is the smallest CS ($\sigma $-CS
), which contains the set $A$.

\begin{defi}
\cite{N7} Let $L$ be a quantum logic and $L_c$ be an $\sigma $-CS in $L$.
Let $f:L\times L_c\to [0,1]$. If the function $f$ fulfill the following
conditions:

\begin{itemize}
\item[(C1)]  for each $a\in L_0$ $f(.,a)$ is a state on $L$;

\item[(C2)]  for each $a\in L_0$ $f(a,a)=1$;

\item[(C3)]  if $\{a_n\}_{n\in\mathcal{A}}\in L_0$, where $\mathcal{A\subset
N}$ and $a_n$ are mutually orthogonal, then for each $b\in L$ 
\[
f(b,\bigvee_{n\in\mathcal{A}}a_n)=\sum_{n\in\mathcal{A}}f(a_n,\bigvee_{n\in%
\mathcal{A}}a_n) f(b,a_n); 
\]
\end{itemize}

then it is called \textit{conditional state}.
\end{defi}

\vskip 1pc

\begin{prop}
\cite{N7} Let $L$ be a q.l.. Let $\{a_i\}_{i=1}^n\in L$, $n\in N$ where $%
a_i\perp a_j$ for $i\neq j$. If for any $i$ there exists a state $\alpha_i$,
such that $\alpha_i(a_i)=1$, then there exists $\sigma $-CS such that for
any $\mathbf{k}=(k_1,k_2,...,k_n)$, where $k_i\in [0;1]$ for $%
i\in\{1,2,...,n\}$ with the property $\sum_{i=1}^n k_i=1$, there exists a
conditional state 
\[
f_{\mathbf{k}}:L\times L_c\to [0;1], 
\]
such that

\begin{enumerate}
\item  for any $i$ and each $d\in L$ $f_{\mathbf{k}}(d,a_i)=\alpha_i(d);$

\item  for each $a_i$ 
\[
f_{\mathbf{k}}(a_i,\bigvee_{i=1}^na_i)=k_i; 
\]
\end{enumerate}
\end{prop}

\vskip 1pc

\begin{defi}
\cite{N7} Let $L$ be a quantum logic and $f$ be a conditional state. Let $%
b\in L$, $a,c\in L_c$ such that $f(c,a)=1$. Then $b$ is independent of $a$
with respect to the state $f(.,c)$ ($b\asymp_{f(.,c)}a$) if $f(b,c)=f(b,a)$.
\end{defi}

\vskip 1pc The classical definition of independency of a probability space $%
(\Omega,\mathcal{B,P)}$ is a special case of this definition, because \vskip %
1pc 
\centerline{$P(A|B)=P(A|\Omega )$ if and only if $ P(A\cap B|\Omega )=
P(A|\Omega )P(B|\Omega )$.} \vskip 2pc

If $L_c$ be CS and $f:L\times L_c\to [0,1]$ is a conditional state, then ( 
\cite{N7})

\begin{itemize}
\item[(i)]  Let $a^\perp,a,c\in L_c$, $b\in L$ and $f(c,a)=f(c,a^\perp )=1$.
Then $b\asymp_{f(.,c)} a$ if and only if $b\asymp_{f(.,c)} a^\perp$.

\item[(ii)]  Let $a,c\in L_c$, $b\in L$ and $f(c,a)=1$. Then $%
b\asymp_{f(.,c)} a$ if and only if $b^\perp \asymp_{f(.,c)} a$.

\item[(iii)]  Let $a,c,b\in L_c$, $b\leftrightarrow a$ and $f(c,a)=f(c,b)=1$%
. Then $b\asymp_{f(.,c)} a$ if and only if $a\asymp_{f(.,c)} b$.
\end{itemize}

\vskip 5pc \centerline {\bf 2. Observables and {\it s-map}} \vskip 1pc

Let $L$ be a quantum logics. Let us denote $L^2 =L\times L$.

\begin{defi2}
Let $L$ be a quantum logic. The map $p:L^2\to [0,1]$ will be called 
\textit{s-map} if the following conditions hold:

\begin{itemize}
\item[(s1)] $p(1,1)=1$;

\item[(s2)]  if there exists $i\in N$ $a\perp b$, then $p(a,b)=0$;

\item[(s3)]  if $a\perp b$, then for any $c\in L$, 
\[
p(a\vee b,c)=p(a,c)+p(b,c) 
\]
\[
p(c,a\vee b)=p(c,a)+p(c,b) 
\]
.
\end{itemize}
\end{defi2}

\vskip 1pc

\begin{prop2}
Let $L$ be a q.l. and let $p$ be a s-map. Let $a,b,c\in L$, then

\begin{itemize}
\item[1.]  if $a\leftrightarrow b$ , then $p(a,b)=p(a\wedge b,a\wedge
b)=p(b,a)$;

\item[2.]  if $a\le b$, then $p(a,b)=p(a,a)$;

\item[3.]  if $a\le b$, then $p(a,c)\le p(b,c)$;

\item[4.]  $p(a,b)\le p(b,b)$;

\item[5.]  if $\nu (b)=p(b,b)$, then $\nu $ is a state on $L$.
\end{itemize}
\end{prop2}

\vskip 1pc

\begin{prop2}
Let $L$ be a q.l., let there be a s-map $p$. Then there exists a conditional
state $f_p$, such that 
\[
p(a,b)=f_p(a,b)f_p(a,1). 
\]

Let $L$ be a q.l. and let $L_c=L-\{0\}$. If $f:L\times L_c\to [0,1]$ is a
conditional state, then there exists a s-map $p_f:L\times L\to [0,1]$.
\end{prop2}

\vskip 1pc

\begin{prop2}
Let $L$ be a quantum logic.

\begin{itemize}
\item[(a)]  If $f$ is a conditional state, then $b\asymp_{f(.,1)} a$ iff $%
p_f(b,a)=p_f(a,a)p_f(b,b)$, where $p_f$ is the s-map generated by $f$.

\item[(b)]  Let $p$ be a s-map. Then $b\asymp_{f_p(.,1)} a$ iff $%
p(b,a)=p(a,a)p(b,b)$, where $f_p$ is the conditional state generated by the
s-map $p$.
\end{itemize}
\end{prop2}

\vskip 1pc Let $\mathcal{B(R)}$ be $\sigma$-algebra of Borel sets. A $\sigma 
$-homomorphism $x:\mathcal{B(R)\to L}$ is called an observable on $L$. If $x$
is an observable, then $R(x):=\{x(E);\quad E\in\mathcal{B(\mathcal{R}) \}}$ is called
range of the observable $x$. It is clear that $R(x)$ is Boolean $\sigma $%
-algebra \cite{V1}. A spectrum of an observable $x$ is defined by the following
way: $\sigma (x)=\cap \{E\in\mathcal{B(R);}$ $x(E)=1\}.$ If $g$ is a real
function, then $g\circ x$ is such observable on $L$ that:

\begin{itemize}
\item[(1.)]  $R(g\circ x)\subset R(x)$;

\item[(2.)]  $\sigma (g\circ x)=\{g(t);\quad t\in \sigma (x)\}$;

\item[(3.)]  for any $E\in\mathcal{B(R)}$

$g\circ x(E) = x(\{t\in \sigma (x); g(t)\in E\})$.
\end{itemize}
We say that $x$ and $y$ are compatible ($x\leftrightarrow y$) if there exist a Boolean 
sub-$\sigma $-algebra  $\mathcal B\subset L$ such that $R(x)\cup R(y)\subset \mathcal B$.
In other words $x\leftrightarrow y$ if for any $E,F\in\mathcal{B}(\mathcal{R})$
$x(E)\leftrightarrow y(F)$.  

We call an observable $x$ a discrete if $\sigma (x)$ is a countable set. It
means, that $\sigma (x)=\{t_i\}_{i\in \mathcal{A}}$, where $\mathcal{A}$$%
\subset \{1,2,...\}$. Let us denote $\mathcal{O}$ the set of all discrete
observables on $L$.

\begin{defi2}
Let $L$ be a $\sigma $-quantum logic and $p:L\times L\to [0;1]$ be a s-map.
Let $x,y$ be some observables on $L$. Then a map $p_{x,y}:\mathcal{%
B(R)\times B(R)\to }$ $[0,1]$, such that 
\[
p_{x,y}(E,F)=p(x(E),y(F)), 
\]
is called a joint distribution for the observables $x$ and $y$ for the $s$-map $p$..
\end{defi2}

\vskip 1pc If $x$ is an observable on $L$ and $m$ is a state on $L$, then $%
m_x(E)=m(x(E))$, $E\in\mathcal{B(R)}$ is a probability distribution for $x$ and 
\[
m(x)=\int_R\lambda m_x(d\lambda ) 
\]
is called the expectation of $x$ in the state $m$, if the integral on the
right side exists. From this follows that for $x\in\mathcal{O}$ 
\[
m(x)=\sum_{t\in\sigma (x)}tm(x(\{t\})) 
\]
and for any real function $g$ we have 
\[
m(g\circ x):=\sum_{t\in\sigma (x)}g(t)m(x(\{t\})). 
\]
\begin{defi2}
Let $L$ be a $\sigma $-quantum logic and $p:L\times L\to [0;1]$ be a s-map.
Let $x,y\in\mathcal{O}$.  Then \[ p(x,y)=\sum_{x_i\in\sigma (x)}\sum_{y_j\in\sigma (y)}x_iy_j
p(x(\{x_i\}),y(\{y_j\})),\]

is called the first joint moment for the observables $x,y$. 
\end{defi2}

 From the previous definition follows that  
\[
p(x,x)=\sum_{x_i\in\sigma (x)}x_i^2p(x(\{x_i\}),x(\{x_i\}))=
\sum_{x_i\in\sigma (x)}x_i^2\nu(x(\{x_i\})):=\nu(g\circ x), 
\]
where $g(t)=t^2$.

Anallogicaly to the classical theory of probability we can 
define notions for example as covariance ($c(.,.)$), variance 
($var(.)$) and correlation coeficient ($r(.,.)$)  
 by the following way:
 \[
c(x,y):=p(x,y)-\nu (x)\nu (y), 
\]

\[var(x):=c(x,x),\]
\[r(x,y)=\frac{c(x,y)}{\sqrt{var(x).var(y)}}.\]

In spite of the classical theory of probability in this case 
$c(x,y)$ is not equal to $c(y,x)$ in generali. 
 \vskip 1pc

\begin{prop2}
Let $L$ be a quantum logic, $p$ be a s-map on $L$ and $\mathcal{O}$ be a set
of all discrete observables on $L$. For each $x,y\in\mathcal{O}$ there exist 
$(\Omega_i,$$\mathcal{S}$$_i,P_i)$ ($i=1,2$) probability spaces and random
variables $\xi_i,\eta_i $ ($i=1,2$), which are $\mathcal{S}$$_i$-measurable
such that:

\begin{itemize}
\item[(a)]  $c(x,y)=cov(\xi_1,\eta_1)$, $c(y,x)=cov(\eta_2,\xi_2)$;

\item[(b)]  $E_i(\xi_i)=\nu(x)$ and $E_i(\eta_i)=\nu (y)$, $i=1,2$;

\item[(c)]  $(c(x,y))^2\le c(x,x)c(y,y)$.
\end{itemize}
\end{prop2}

Proof. If $x,y\in\mathcal{O}$, then $\sigma (x)=\{x_k\}_{k\in \mathcal{A}}$, 
$\sigma (y)=\{y_j\}_{j\in\mathcal{D}}$, where $\mathcal{A}$,$\mathcal{D}$ $%
\subset N$. Let us denote $\Omega_1=\{(x_k,y_j);\quad k\in\mathcal{A}$ , $%
j\in\mathcal{D}$$\}$, $\Omega_2=\{(y_j,x_k);\quad k\in\mathcal{A}$ , $j\in%
\mathcal{D}$$\}$ and $\mathcal{S}$$_i=2^{\Omega_i}$, $i=1,2$. Then $%
(\Omega_i,\mathcal{S}$$_i)$ is the measurable space.

If $p$ is a s-map, then from the properties of $p$ follows, that $P_1=p_{x,y}
$ is the probability measure on $(\Omega_1,\mathcal{S}$$_1)$ and $P_2=p_{y,x}
$ is the probability measure on the measurable space $(\Omega_2,\mathcal{S}$$%
_2)$. From this 
\[
E_i(\xi_i)=\sum_{k}x_kP_i(\xi_1=x_k)=\nu (x). 
\]
Anallogicaly we get 
\[
E_i(\eta_i)=\nu (y). 
\]

(b) Let us denote 
\[
\xi_1((x_k,y_j))=x_k,\quad \eta_1((x_k,y_j))=y_j, 
\]
\[
\xi_2((y_j,x_k))=x_k,\quad \eta_2((y_j,x_k))=y_j. 
\]
It is clear, that 
\[
P_1(\{\omega\in\Omega_1;\quad \xi_1(\omega)=x_k\})= P_1\{(x_k,y_j);j\in%
\mathcal{D\}} 
\]
From this follows, that 
\[
P_1(\{\omega\in\Omega_1;\quad \xi_1(\omega)=x_k\})= p(x(\{x_k\}),y(\sigma
(y)))=p(x(\{x_k\},1)=\nu (x(\{x_k\})). 
\]
Anallogicaly from this follows, that 
\[
P_i(\xi_i=x_k)=\nu (x(\{x_k\}) 
\]
and 
\[
P_i(\eta_i=y_j)=\nu (y(\{y_j\})) 
\]

(a) From the theory of probability follows, that 
\[
var (\xi_1,\eta_1 )= \sum_{k}\sum_{j}(x_k-\nu (x))(y_j-\nu
(y))P_1(\xi_1=x_k,\eta_1=y_j) ) 
\]
Let us denote $a_k=x(\{x_k\})$ and $b_j=y(\{y_j\})$. Then 
\[
cov(\xi_1,\eta_1)=\sum_k(x_k-\nu (x))(y_j-\nu (y))p(a_k,b_j) 
\]
and 
\[
cov(\xi_2,\eta_2)=\sum_k(x_k-\nu (x))(y_j-\nu (y))p(b_j,a_k). 
\]
Because 
\[
\sum_{k,j}p(a_k,b_j)=\sum_{k,j}p(b_j,a_k)=1, 
\]
\[
\sum_kp(a_k,b_j)=\sum_kp(b_j,a_k)=\nu (b_j) 
\]
and 
\[
\sum_jp(a_k,b_j)=\sum_jp(b_j,a_k)=\nu (a_k) 
\]
we have 
\[
\sum_k(x_k-\nu (x))(y_j-\nu (y))p(b_j,a_k)=p(x,y)-\nu (x)\nu (y)=c(x,y). 
\]
Anallogically 
\[
var (\xi_2,\eta_2)=p(y,x)-\nu (x)\nu (y)=c(y,x). 
\]

(c) Because $(cov(\xi _{i},\eta _{i}))^{2}\leq cov(\xi _{i},\xi
_{i})cov(\eta _{i},\eta _{i})$ and $cov(\xi _{i},\xi _{i})=c(x,x)$, $%
cov(\eta _{i},\eta _{i})=c(y,y)$ we have 
\[
(c(x,y))^{2}\leq c(x,x)c(y,y), 
\]
\[
(c(y,x))^{2}\leq c(x,x)c(y,y). 
\]
\vskip1pc (Q.E.D.)

\bigskip

\vskip 2pc
\begin{prop2}
Let $L$ be a quantum logic and $\cal O$ be the set of all discrete observables 
on $L$. Let $x,y\in\mathcal{O}$. Then
\begin{itemize}
\item[(i)] $c(x,y)=p(g_x\circ x,g_y\circ y)$,  
where $g_z$ is a real 
function such that $g_z(t)=t-\nu(z)$, for $z\in \mathcal{O}$,
;
\item[(ii)] $r(x,y)\in [-1,1]$;
\item[(iii)] if $x\leftrightarrow y$, then $c(x,y)=c(y,x)$ and $r(x,y)=r(y,x)$.
\end{itemize}
\end{prop2}
Proof. Let $x,y\in \mathcal{O}$. Then $\sigma (x)=\{x_i\}_{i\in\mathcal{A}}$
 and $\sigma (y)=\{y_j\}_{j\in\mathcal{B}}$, where $\mathcal{A},\mathcal{B}$ are
countabte sets of indexes.

(i) From the definition $g_x\circ x,g_y\circ y$ we have
$$\begin{array}{clcr}p(g_x\circ x,g_y\circ y)&=\sum_{x_i\in\sigma (x)}
\sum_{y_j\in\sigma (y)}g_x(x_i)g_y(y_j)p(x(x_i),y(y_j))\\
&=\sum_{x_i\in\sigma (x)}\sum_{y_j\in\sigma (y)} (x_i-\nu (x))(y_j-\nu (y))
p(x(x_i),y(y_j))\\
&=\sum_{x_i\in\sigma (x)}\sum_{y_j\in\sigma (y)} x_iy_jp(x(x_i),y(y_j)) \\
&-\sum_{x_i\in\sigma (x)}\sum_{y_j\in\sigma (y)} \nu (x)\nu (y)p(x(x_i),y(y_j))\\
&=p(x,y)-\nu (x) \nu(y)=c(x,y).\end{array}$$

(ii) From the previous proposition we know that there exist two 
probability spaces $(\Omega_k,\mathcal{F}_k,P_k)$ and random variables 
$\xi_k,\eta_k$, $k=1,2$ such that $c(x,y)=cov(\xi_1,\eta_1)$ and 
$c(x,y)=cov(\xi_2,\eta_2)$. Moreover $c(x,x)=cov(\xi_k,\xi_k)=var(\xi_k)$.
Anallogicaly $c(y,y)=cov(\eta_k,\eta_k)=var(\eta_k)$. Then
$$r(x,y)=\frac{c(x,y)}{\sqrt{c(x,x)c(y,y)}}=\frac{cov(\xi_1,\eta_1)}
{\sqrt{var(\xi_1)var(\eta_1)}}=\rho (\xi_1,\eta_1)=\rho_1$$ 
and
$$r(y,x)=\frac{c(y,x)}{\sqrt{c(x,x)c(y,y)}}=\frac{cov(\xi_2,\eta_2)}
{\sqrt{var(\xi_2)var(\eta_2)}}=\rho (\xi_2,\eta_2)=\rho_2.$$
becuase $\rho_k$  is correlation 
coeficient on the probability space   $(\Omega_k,\mathcal{F}_k,P_k)$, 
 then $\rho_k\in [-1,1]$ for for $k=1,2$. 
 From this follows that $r(x,y),r(y,x)\in [-1,1]$. 
Also this fact follows  immidiately from the Proposition 2.4 (c).

(iii) Let $x\leftrightarrow y$. Then for each $x_i\in\sigma (x)$ and each 
 for  $y_j\in\sigma (y)$ we have
$$p(x(x_i),y(y_j))=p(y(y_j),x(x_i))$$
and so
$$\begin{array}{clcr}p(x,y)&=
\sum_{x_i\in\sigma (x)}\sum_{y_j\in\sigma (y)} x_iy_jp(x(x_i),y(y_j))\\
&=\sum_{x_i\in\sigma (x)}\sum_{y_j\in\sigma (y)} x_iy_jp(y(y_j),x(x_i))\\
&=p(y,x).\end{array}$$
From this follows that 
$$c(x,y)=c(y,x).$$
(Q.E.D.)
\begin{ex2}.
Let  $L=\{a,a^\perp ,b,b^\perp , 0,1\}$. Let  
$c\vee d=1$ if $c\neq d$ and $c,d\in L-\{0\}$. Let $c\wedge d=0$ if 
$c\neq d$ and $c,d\in L-\{1\}$. Let $(d^\perp)^\perp =d$  for $d\in L$ 
and $1^\perp =0$.
   It is clear that
$L$ is a quantum logic and $B_d=\{d,d^\perp ,0,1\}$, $d\in \{a,b\}$ is Booelan 
algebra. Let $f(s,t)$ is defined by the following way:

$$
\vbox{\offinterlineskip
\halign{
\strut\vrule $#$ & \vrule $#$ &\vrule $#$
&\vrule $#$ &\vrule $#$&\vrule $#$\hfil\vrule \cr
\noalign{\hrule}
\quad s/t   \quad &\quad a\quad &\quad a^\perp\quad &\quad b   \quad &\quad b^\perp \quad&\quad 1  \quad\cr
\noalign{\hrule}
\quad a     \quad &\quad 1\quad &\quad 0 \quad &\quad 0.4 \quad &\quad 0.4    \quad &\quad 0.4\quad\cr
\noalign{\hrule}
\quad a^\perp\quad &\quad 0\quad &\quad 1      \quad &\quad 0.6\quad & \quad 0.6   \quad &\quad 0.6\quad\cr
\noalign{\hrule}
\quad b\quad &\quad 0.2\quad &\quad11/30\quad &\quad 1\quad &\quad 0\quad &\quad 0.3\quad\cr
\noalign{\hrule}
\quad b^\perp\quad &\quad 0.8\quad &\quad 19/30\quad &\quad 0\quad  & \quad 1\quad&\quad 0.7\quad \cr
\noalign{\hrule}
}}$$
From $f$ this we can compute $p_f(s,t)$ . Then we get:
$$
\vbox{\offinterlineskip
\halign{
\strut\vrule $#$ & \vrule $#$ &\vrule $#$
&\vrule $#$ &\vrule $#$\hfil\vrule \cr
\noalign{\hrule}
\quad s/t \quad &\quad a\quad &\quad a^\perp\quad &\quad b   \quad &\quad b^\perp \quad\cr
\noalign{\hrule}
\quad a\quad &\quad 0.4 \quad &\quad 0\quad &\quad 0.12 \quad &\quad 0.28   \quad \cr
\noalign{\hrule}
\quad a^\perp\quad &\quad 0\quad &\quad 0.6\quad &\quad 0.18 \quad & \quad 0.42 \quad \cr
\noalign{\hrule}
\quad b\quad &\quad 0.08\quad &\quad 0.22\quad &\quad 0.3\quad &\quad 0\quad \cr
\noalign{\hrule}
\quad b^\perp\quad &\quad 0.32\quad &\quad 0.38\quad &\quad 0\quad  & \quad 0.7\quad \cr
\noalign{\hrule}
}}$$
We can see that $p_f(a,b)=p_f(a,a)p_f(b,b)$, but $p_f(b,a)\ne p_f(b,b)p_f(a,a)$.
In the following we will write $p_f=p$.
Let $x,y $ be  observales
on $L$ such that $R(x)=\{a,a^\perp ,0,1\}=B_a$, and $R(y)= \{b,b^\perp ,0,1\}=B_b$. It is easy to 
see, that $x$ is not compatible with $y$. Let, for example,
$$x(-1)=a\quad x(1)=a^\bot$$
$$y(0)=b\quad y(5)=b^\bot. $$
In the following tables we have the joint distributions $p_{x,y}$ and $p_{y,x}$. 

$$
\vbox{\offinterlineskip
\halign{
\strut\vrule $#$ & \vrule $#$ &\vrule $#$\hfil\vrule \cr
\noalign{\hrule}
\quad p_{x,y} \quad &\quad 0\quad &\quad 5\quad \cr
\noalign{\hrule}
\quad -1\quad &\quad 0.12 \quad &\quad 0.28\quad \cr
\noalign{\hrule}
\quad 1\quad &\quad 0.18 \quad & \quad 0.42 \quad \cr
\noalign{\hrule}
}}$$
$$
\vbox{\offinterlineskip
\halign{
\strut\vrule $#$ & \vrule $#$ &\vrule $#$\hfil\vrule \cr
\noalign{\hrule}
\quad p_{y,x} \quad &\quad -1\quad &\quad 1\quad \cr
\noalign{\hrule}
\quad 0\quad &\quad 0.08 \quad &\quad 0.22\quad \cr
\noalign{\hrule}
\quad 5\quad &\quad 0.32 \quad & \quad 0.38 \quad \cr
\noalign{\hrule}
}}$$
Now we can compute the following characteristics:
$$\nu (x)=-1\times  0.4+0.2\times 0.6=0.2,$$
$$\nu (y)=0\times 0.3+5\times 0.7=3.5,$$  
$$p(x,y)=-5\times 0.28+5\times 0.42=0.7,$$
$$p(y,x)=-5\times 0.32+5\times 0.38=0.3,$$
$$c(x,y)=p(x,y)-\nu (x)\nu (y)=0.7-0.2\times 3.5=0,$$
$$c(y,x)=p(y,x)-\nu (x)\nu (y)=0.3-0.2\times 3.5=-0.4,$$
$$c(x,x)=0.96\quad\quad\quad c(y,y)=5.25,$$
$$r(x,y)=0 \quad\quad r(y,x)\doteq 0.178.$$
In the end we can write these results to "the  covariance matrix":
\[ \left( \begin{array}{cc}
c(x,x)& c(x,y)  \\
c(y,x) & c(y,y) \\
\end{array} \right)= \left( \begin{array}{cc}
0.96 & 0  \\
0.178 & 5.25 \\
\end{array} 
\right)\]
We see that "the covariance matrix" need not to be  symmetry. In the classical 
theory of probability, where we suppose that all random variables are 
compatible, have to be symmetry.

\end{ex2} 
\vskip 2pc
\begin{ex2}.
Let  $L$ the same quantum logic as in the Example 2.1.
 Let $p(s,t)$ is defined by the following way:

$$
\vbox{\offinterlineskip
\halign{
\strut\vrule $#$ & \vrule $#$ &\vrule $#$
&\vrule $#$ &\vrule $#$\hfil\vrule \cr
\noalign{\hrule}
\quad s/t \quad &\quad a\quad &\quad a^\perp\quad &\quad b   \quad &\quad b^\perp \quad\cr
\noalign{\hrule}
\quad a\quad &\quad 0.4 \quad &\quad 0\quad &\quad 0.08 \quad &\quad 0.38  
\quad \cr
\noalign{\hrule}
\quad a^\perp\quad &\quad 0\quad &\quad 0.6\quad &\quad 0.22 \quad & \quad 0.32
 \quad \cr
\noalign{\hrule}
\quad b\quad &\quad 0.08\quad &\quad 0.22\quad &\quad 0.3\quad &\quad 0
\quad \cr
\noalign{\hrule}
\quad b^\perp\quad &\quad 0.32\quad &\quad 0.38\quad &\quad 0\quad  & 
\quad 0.7\quad \cr
\noalign{\hrule}
}}$$

Let $x,y $ be  observables
on $L$ such that $R(x)=\{a,a^\perp ,0,1\}$, and $R(y)= \{b,b^\perp ,0,1\}$. 
It is easy to 
see, that $x$ is not compatible with $y$. Let, for example,
$$x(-1)=a\quad x(1)=a^\bot$$
$$y(0)=b\quad y(5)=b^\bot. $$
In this case  $p_{x,y} =p_{y,x}$. 
$$
\vbox{\offinterlineskip
\halign{
\strut\vrule $#$ & \vrule $#$ &\vrule $#$\hfil\vrule \cr
\noalign{\hrule}
\quad p_{x,y} \quad &\quad 0\quad &\quad 5\quad \cr
\noalign{\hrule}
\quad -1\quad &\quad 0.08 \quad &\quad 0.38\quad \cr
\noalign{\hrule}
\quad 1\quad &\quad 0.22 \quad & \quad 0.32 \quad \cr
\noalign{\hrule}
}}$$

Now we get 
$$\nu (x)=0.2,\quad\quad\nu (y)=3.5,$$  
$$p(x,y)=p(y,x)=-0.3,\quad\quad c(x,y)=c(y,x)=-0.4,$$
$$c(x,x)=0.96,\quad\quad c(y,y)=5.25,$$
$$r(x,y)=r(y,x)\doteq 0.178.$$
In the end we can write these results to the  covariance matrix:
\[\left( \begin{array}{cc}
c(x,x)& c(x,y)  \\
c(y,x) & c(y,y) \\
\end{array} \right)= \left( \begin{array}{cc}
0.96 & 0.178  \\
0.178 & 5.25 \\
\end{array} 
\right)\]
We see that the covariance matrix is  symmetry as  in the classical 
theory of probability, but $x,y$ are not compatible.

\end{ex2}
\vskip 2pc
 In the end we can say that we cannot prove that two observables are 
compatible by using statistics but we can prove that they are not 
compatible.


\begin{thebibliography}{99}

\bibitem{A} Accardi  L., The probabilistic roots of the quantum mechanical paradoxes.
{\em The wave--particle dualism.  A tribute to Louis de Broglie on his 90th 
Birthday,} ed. S. Diner, D. Fargue, G. Lochak and F. Selleri
(D. Reidel Publ. Company, Dordrecht, 297--330, 1984).

\bibitem{Bal2} Ballentine L. E. ,  Probability theory in quantum mechanics. {\it American
J. of Physics}, {\bf 54,} 883-888 (1986).

\bibitem{Bal1} Ballentine L. E.,  Interpretations of probability and quantum theory.
Proc. Conf. {\it Foundations of Probability and Physics,} ed. A. Khrennikov.
{\it Q. Prob. White Noise Anal.}, {\bf 13}, 71-84, WSP, Singapore (2001).


\bibitem{Gud} Gudder S. P., An approach to quantum probability. Proc. Conf.
{\it Foundations of Probability and Physics,} ed. A. Khrennikov.
Quantum Prob. White Noise Anal., {\bf 13}, 147-160, WSP, Singapore (2001).

\bibitem{DEM} De Muynck W. M., Interpretations of quantum mechanics, and interpretations of 
violation of Bell's inequality. 
Proc. Conf. {\it Foundations of Probability and Physics,} ed. A. Khrennikov.
{\it Q. Prob. White Noise Anal.}, {\bf 13}, 95-114, WSP, Singapore (2001).


\bibitem{KHR1}  Khrennikov A. Yu., {\it Contextual viewpoint to quantum stochastics.} hep-th/0112076.

\bibitem{KHR2}  Khrennikov A. Yu., {\it Interpretations of Probability},
VSP Int. Sc. Publ., Utrecht (1999).

\bibitem{KHR3} Khrennikov A. Yu., Linear representations of probabilistic transformations induced by
context transitions. {\it J. Phys.A: Math. Gen.,} {\bf 34}, 9965-9981 (2001).

\bibitem{KHR4} Khrennikov A. Yu., {\it Hyperbolic quantum mechanics.} quant-ph/0101002 (2000).

\bibitem{KHR5} Khrennikov A. Yu., Contextualist viewpoint to 
Greenberger-Horne-Zeilinger paradox. {\it Phys. Lett.}, A, {\bf 278}, 307-314 (2001).

\bibitem{KHR6} Khrennikov A. Yu., Frequency analysis of the EPR-Bell argumentation. {\it Foundations of Physics,}
{\bf 32,} 1159-1174 (2002).


\bibitem{K1}  Kolmogoroff A.N., Grundbegriffe der
Wahrscheikch\-keit\-sre\-chnung, Springer, Berlin, (1933)

\bibitem{K2} Kolmogorov  A. N., {\it The Theory of Probability.} In: A.D.Alexandrov,
A.N.Kol\-mo\-go\-rov, M.A.Lavrent'ev (Eds.) {\it Mathematics, Its Content, Methods,
and Meaning,} {\bf 2,} M.I.T. Press  (1965).

\bibitem{Ren} Renyi A,  On a new axiomatics of probability theory. {\it Acta Mat. Acad. Sc. Hung.},
{\bf 6}, 285-335 (1955).

\bibitem{Cox} Cox R., {\it The algebra of probable inference.} J. Hopkins Univ. Press, 
Baltimore (1961).


\bibitem{Bel} Beltrametti  E. and Cassinelli G., {\it  The logic of quantum mechanics.}
Addison-Wesley, Reading, Mass (1981).


\bibitem{D2}  Dvure\v censkij A., Pulmannov\'a S., New Trends in Quantum
Structures, Kluwer Acad. Publ. (2000).



\bibitem{N2}  N\'an\'asiov\'a O., On conditional probabilities on quantum
logic. {\it Int. J. Theor. Phys.}, {\bf 25}, 155 - 162 (1987).

\bibitem{N3}  N\'an\'asiov\'a O., Orderinng of observables and
characterization of conditional expectation on a quantum logic. {\it Math.
Slovaca,} {\bf 37,}  323-340 (1987).

\bibitem{N4}  N\'an\'asiov\'a O., Observables and expectation on the Pt\'ak
sum. {\it Tatra Mount. Publ. Math.,} {\bf 3}, 65-76  (1993).

\bibitem{N5}  N\'an\'asiov\'a O., States and homomorphism on the Pt\'ak sum.
{\it Int. J. Theor. Pfys.,} {\bf 32}, 1957-1964  (1993).

\bibitem{N6}  N\'an\'asiov\'a O., A note on the independent events on
aquantum logic. {\it  Busefal,}  {\bf 76,}  53-57 (1998).

\bibitem{N7}  N\'an\'asiov\'a O., Principle conditioning. Preprint (2001).

\bibitem{P1}  Pt\'ak P., Pulmannov\'a S., {\it Quantum Logics.} Kluwer Acad.
Press, Bratislava (1991).

\bibitem{R1}  Rie\v can B., Neubrun T., {\it Measure theory.}



\bibitem{V1}  Varadarajan V., Geometry of quantum theory, Princeton, New
Jersey, D. Van Nostrand, (1968)
\end{thebibliography}
\end{document}